\documentclass[structabstract]{aa}

\usepackage{graphicx,epsfig}
\usepackage[dvips]{rotating}
\usepackage{amsmath}
\usepackage{txfonts}
\usepackage{multirow}

\usepackage[colorlinks=true,linkcolor=black,citecolor=black]{hyperref}

\usepackage{natbib}
\bibpunct{(}{)}{;}{a}{}{,}

\newcommand{\ergcm}[1]{$10^{#1}$ erg cm$^{-2}$ s$^{-1}$}

\newcommand{\hcm}[1]{$\times\, 10^{#1}$ cm$^{-2}$}

\newcommand{\ltsima}{$\buildrel < \over \sim$}
\newcommand{\lsim}{\lower.5ex\hbox{\ltsima}}
\newcommand{\gtsima}{$\buildrel > \over \sim$}
\newcommand{\gsim}{\lower.5ex\hbox{\gtsima}}

\newcommand{\xmm}{XMM-Newton}

\newcommand{\chandra}{\textit{Chandra}}

\newcommand{\hess}{H.E.S.S.}
\newcommand{\gon}{G0.9$+$0.1}
\newcommand{\psrj}{PSR\,J1747$-$2809}

\newcommand{\cxou}{CXOU J174722.8$-$280915}

\newcommand{\gr}{\ensuremath{\gamma}-ray}
\newcommand{\eemin}{E_{\mathrm{e,min}}}
\newcommand{\eemax}{E_{\mathrm{e,max}}}

\newcommand{\degree}[1]{${#1}^{\circ}$}

\newcommand{\scenone}{\textit{Scenario I}}
\newcommand{\scentwo}{\textit{Scenario II}}

\begin{document}

\title{Spatially resolved X-ray spectroscopy and modeling of the nonthermal emission of the PWN in \gon}

\author{M. Holler\inst{1,2}
\and
F.\,M. Sch\"ock\inst{1}
\and
P. Eger\inst{1}
\and
D. Kie\ss ling\inst{1}
\and
K. Valerius\inst{1}
\and
C. Stegmann\inst{2,3,4}
}

\institute{
Erlangen Centre for Astroparticle Physics, Erwin-Rommel-Stra\ss e 1, 91058 Erlangen, Germany
\and{Institut f\"{u}r Physik und Astronomie, Universit\"{a}t Potsdam, 14476 Potsdam-Golm, Germany}
\and{DESY, Platanenallee 6, 15738 Zeuthen, Germany}
\and{Formerly at $^{1}$}
}

\offprints{Markus\ ~Holler, \email{markus.holler@desy.de}}

\titlerunning{Spatially resolved X-ray spectroscopy and modeling of the PWN in \gon}
\authorrunning{M. Holler et al.}

\date{Received 19 September 2011 / Accepted 12 January 2012}

\abstract{}{We performed a spatially resolved spectral X-ray study of the pulsar wind nebula (PWN) in the supernova remnant \gon. Furthermore, we modeled its nonthermal emission in the X-ray and very high-energy (VHE, $E>100\,\mathrm{GeV}$) $\gamma$-ray regime.}{Using \chandra\ ACIS-S3 data, we investigated the east-west dependence of the spectral properties of \gon\ by calculating hardness ratios. We analyzed the EPIC-MOS and EPIC-pn data of two on-axis observations of the \xmm\ telescope and extracted spectra of four annulus-shaped regions, centered on the region of brightest emission of the source. A radially symmetric leptonic model was applied in order to reproduce the observed X-ray emission of the inner part of the PWN. Using the optimized model parameter values obtained from the X-ray analysis, we then compared the modeled inverse Compton (IC) radiation with the published \hess\ \gr\ data.}{The spectral index within the four annuli increases with growing distance to the pulsar, whereas the surface brightness drops. With the adopted model we are able to reproduce the characteristics of the X-ray spectra. The model results for the VHE $\gamma$ radiation, however, strongly deviate from the \hess\ data.}{}

\maketitle{}

\section{Introduction}
\label{introduction}

The composite supernova remnant (SNR) \gon\ is located less than one degree from the Galactic center (GC). It was discovered with the Molonglo radio telescope by \citet{Kesteven1968}. \citet{Helfand1987} found the overall radio morphology to be dominated by a luminous core with a diameter of $\approx 2\,\arcmin$ surrounded by a fainter, but still detectable shell (diameter $\approx 8\,\arcmin$). 

The first firm detection of \gon\ in X-rays has been achieved with the BeppoSAX satellite \citep{Mereghetti1998}. Follow-up observations with \chandra\ \citep{Gaensler2001} and \xmm\ \citep{Porquet2003} unambiguously identified the core of the SNR as a pulsar wind nebula (PWN). The morphology of \gon\ seen in the \chandra\ data set reveals an axial symmetry where the symmetry axis lies at an angle of $\approx\,$\degree{165}, measured counterclockwise from north \citep{Gaensler2001}. An unresolved source (\cxou) is found along the symmetry axis. Due to the high sensitivity of \xmm, \citet{Porquet2003} were also able to detect faint X-ray emission from the shell of the SNR. Comparing different regions in the PWN, they observed a positive gradient of the spectral index from east to west. This is interpreted as fast rotation with resulting Doppler boosting of the leptons that constitute the PWN, requiring an anomalously large extent of the termination shock with a shock radius of $R_{\mathrm{S}}>1\,\mathrm{pc}$ as suggested by \citet{Gaensler2001}. 

\citet{Dubner2008} have compared newer radio observations with a revised \xmm\ analysis. These authors show that the X-ray PWN almost fills the size of the radio core, which indicates a moderate magnetic field \citep[see e.g.][]{Gaensler2006}. They support the axial symmetry suggested by \citet{Gaensler2001}, but do not see any spectral variations in the radio regime, contrary to the ones in X-rays reported by \citet{Porquet2003}. The authors did not find a radio counterpart of \cxou.      

\gon\ has been detected with \hess\ \citep{Aharonian2005} in very high-energy (VHE) \gr s above $200\,$GeV. With a flux of only $2\,\%$ compared to the one of the Crab Nebula, \gon\ was the faintest known VHE $\gamma$-ray source at that time. However, due to the large field of view of \hess, the telescopes simultaneously observed \gon\ when pointing towards the GC, resulting in a high significance of $13\, \sigma$ due to the long exposure. For \hess, the source appears point-like, with an upper limit on the intrinsic angular extent between $1.3\, \arcmin$ assuming a Gaussian emission region and $2.2\, \arcmin$ for the emission from a thin shell (both at $95\,\%$ confidence level).

In \citeyear{Camilo2009}, \citeauthor{Camilo2009} discovered \psrj, the pulsar powering \gon, using the NRAO Green Bank Telescope at a frequency of $2\,$GHz. The authors report a strongly scattered and dispersed signal with a pulsation period of $P = 52\,\mathrm{ms}$. The characteristic age of the pulsar \citep{Manchester1977} was derived as $\tau_c = 5.3\, \mathrm{kyr}$ and the spin-down luminosity as $\dot{E} = 4.3 \times 10^{37}\,\mathrm{erg}/\mathrm{s}$. Notably, this corresponds to the second-highest known spin-down luminosity of Galactic pulsars, only surpassed by the Crab pulsar. \citet{Camilo2009} estimate the true age of \psrj\ at $2-3\,\mathrm{kyr}$, hence well below the characteristic age. The high scattering of the signal increases the positional uncertainty to about $3\,\arcmin$, neither proving nor disproving that \cxou\ is the X-ray counterpart of \psrj. Due to the high absorption and hard spectrum, \citet{Gaensler2001} ruled out the possibility that \cxou\ corresponds to a foreground star. The authors estimate that the probability of its emission originating in a background active galactic nucleus is $\approx 10^{-2}$. 

Up to the present, the distance to \gon\ is not very well constrained. Many authors \citep[for example,][]{Aharonian2005,Dubner2008} adopt $d = 8.5\, \mathrm{kpc}$, which would imply \gon\ lying in the direct vicinity of the GC. Applying the NE2001 electron model \citep{Cordes2002} to the dispersion measure of \psrj, \citet{Camilo2009} obtain $d \approx 13\, \mathrm{kpc}$, but admit that this model could be substantially in error for sources toward the inner Galactic regions. Due to this uncertainty, they propose an approximate distance of $8-16\,\mathrm{kpc}$, leaving this problem unsolved.

In the subsequent section, we describe the analysis of the \xmm\ observations used for this work. The following part contains a detailed study of the spectral properties of the PWN in \gon\ in X-rays, using data from \chandra , as well as from \xmm . In Section~\ref{section_model} we outline the radially symmetric leptonic model applied to this source. The results of the modeling for two different lepton injection spectra are presented and compared in Section~\ref{section_results}. Using the parameters optimized to fit the X-ray data, we calculate the inverse Compton (IC) emission and compare it with the \hess\ results in Section~\ref{section_tev}.
      
\section{\xmm\ observations and data analysis}
\label{section_observations}

\gon\ was observed on-axis twice by the \xmm\ telescope \citep{Jansen2001} in 2000 and 2003. 
\begin{table}
\caption{Analyzed \gon\ \xmm\ observations.}
\label{tab_obs}
\centering
\begin{tabular}{ccccc}
\hline\hline\noalign{\smallskip}
\multicolumn{1}{c}{Observation} &
\multicolumn{1}{c}{Year} &
\multicolumn{1}{c}{Instrument} &
\multicolumn{2}{c}{Exposures (ks)} \\
\multicolumn{1}{c}{ID} & &
\multicolumn{1}{c}{EPIC} &
\multicolumn{1}{c}{performed$^{(1)}$}& 
\multicolumn{1}{c}{net$^{(2)}$} \\
\noalign{\smallskip}\hline\noalign{\smallskip}
\multirow{2}{*}{0112970201} & 2000 & pn & 11.7 & 11.3 \\
 & 2000 & MOS1/2 & 17.2 & 17.2 \\ 
\noalign{\smallskip}\hline\noalign{\smallskip}
\multirow{3}{*}{0144220101} & 2003 & pn & 43.7 & 26.9 \\
& 2003 & MOS1 & 49.4 & 44.1 \\
& 2003 & MOS2 & 49.4 & 45.3 \\
\noalign{\smallskip}\hline\noalign{\smallskip}
\end{tabular}
\\
$^{(1)}$ Before background screening\\
$^{(2)}$ After background screening
\end{table}
Table~\ref{tab_obs} lists the properties of these two observations used in our analysis. There is also a third data set available (Observation ID (OID): 0205240101), but the high off-axis position of approximately $12\,\arcmin$ makes a spatially resolved spectral analysis difficult due to the strong distortion of the point spread function (PSF) at large off-axis angles. The earlier data set (OID: 0112970201) has already been extensively analyzed by \citet{Porquet2003}.

During both on-axis observations, the medium filter was used for the EPIC-MOS \citep{Turner2001} and EPIC-pn \citep{Strueder2001} cameras. The MOS cameras were operated in the standard full-frame mode in both cases, whereas the pn camera was operated in the extended full-frame mode (2000) and in the large window mode (2003).  

We analyzed the data using version 9.0.0 of the Science Analysis System (SAS) provided by the \xmm\ Science Operations Centre, together with tools from the FTOOLS package \citep{Blackburn1995}. To reduce the influence of soft proton flaring, we selected good time intervals by applying a maximum count threshold to the high-energy lightcurve ($7-15\,$keV) obtained with the standard analysis chain. Values of eight, respectively two, background counts per second were chosen for the pn and for the two MOS cameras. The resulting net exposures are also shown in Table~\ref{tab_obs}. For each data set, we only selected good (FLAG $= 0$) single and multiple (PATTERN $\leq 4$ for pn and $\leq 12$ for MOS) events. Since the source is extended and has a moderate surface brightness, it is not necessary to correct for pile-up. 

For each camera and observation, we defined an infield background region located near the source and on the same CCD chip. For extended sources with low surface brightness like \gon, a proper treatment of all background components is particularly important. Therefore we used the method for particle background subtraction as described by \citet{Majerowicz2002} which was applied to the infield background, as well as to the respective source region. To take the vignetting of the telescopes into account, we adopted the weighting method introduced by \citet{Arnaud2001}.   

Channels were grouped to a signal-to-noise ratio of five for all spectra. 

\section{Spectral properties}
\label{section_spectra}

Due to its small point spread function and low background level, \chandra\ is especially well suited to investigating spectral differences on a small morphological scale. Using the $35\,$ks \chandra\ data set (OID: 1036) previously analyzed by \citet{Gaensler2001}, we calculated the hardness ratios using the counts inside two opposite annulus halves with inner and outer radii $5\,\arcsec$ and $45\,\arcsec$, respectively.  
\begin{figure}
 \centering
 \includegraphics[clip,width=\linewidth]{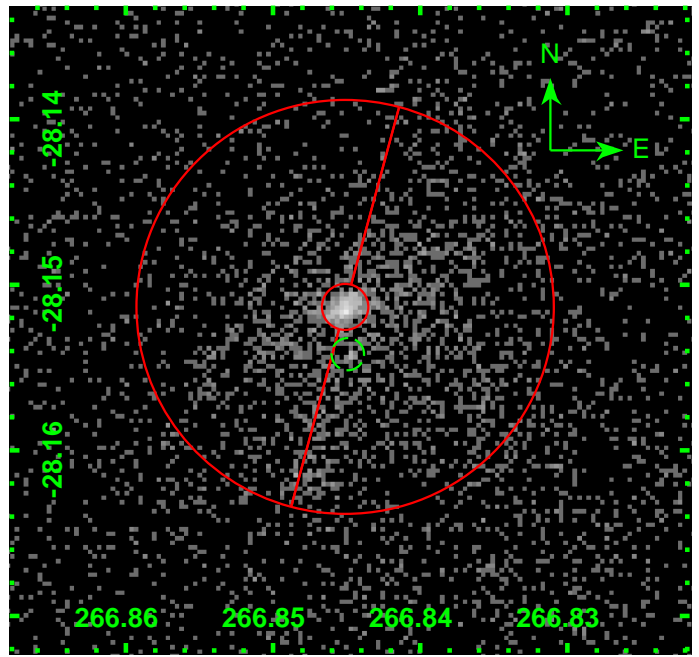}
 \caption{\chandra\ count map in the energy range $3-8\,$keV. Shown in red are the two annulus halves chosen for the hardness ratio calculation (centered on the region of brightest emission in this case). The point source \cxou\ is encompassed by a dashed green circle.}
 \label{fig_chandra}
\end{figure}
The halves are separated by the aforementioned symmetry axis, as illustrated in Fig.~\ref{fig_chandra}. They were centered once on \cxou\ and once on the region of brightest emission. We calculated the fractional difference using the \textit{Bayesian Estimation of Hardness Ratios} from \citet{Park2006}:
\begin{equation}
\mathrm{HR}=\frac{\lambda_{\mathrm{H}}-\lambda_{\mathrm{S}}}{\lambda_{\mathrm{H}}+\lambda_{\mathrm{S}}},
\end{equation}
where $\lambda_{\mathrm{S}}$ and $\lambda_{\mathrm{H}}$ are the expected soft and hard source count intensities, calculated from the measured ones by taking the different effective and extraction areas into account. The low and high energy bands are $3-5\,$keV and $5-8\,$keV, respectively. 
\begin{table}
\caption{Hardness ratios of the estimated source count intensities within the annulus halves. The errors are given at $68\,\%$ confidence level.}
\label{tab_behr}
\centering
\begin{tabular}{ccc}
\hline\hline\noalign{\smallskip}
Center & Orientation & HR \\
\noalign{\smallskip}\hline\noalign{\smallskip}
\multirow{2}{*}{\cxou} & East & $-0.37^{+0.04}_{-0.05}$ \\ \noalign{\smallskip}
& West & $-0.25^{+0.04}_{-0.04}$ \\
\noalign{\smallskip}\hline\noalign{\smallskip}
\multirow{2}{*}{Brightest emission} & East & $-0.36^{+0.04}_{-0.04}$ \\ \noalign{\smallskip}
 & West & $-0.29^{+0.06}_{-0.06}$ \\ 
\noalign{\smallskip}\hline\noalign{\smallskip}
\end{tabular}
\end{table}
The obtained values are given in Table~\ref{tab_behr}. When centering on \cxou , the resulting hardness ratios of the eastern and western halves are inconsistent. The results are, however, comparable within the statistical errors when the halves are centered on the region of brightest emission.

For the spectral modeling of the \xmm\ data, we used version 12.7.0 of the XSPEC tool \citep{Arnaud1996}. The extracted spectra of the pn and MOS cameras of the 2000 and 2003 observations were fitted in parallel with an absorbed power-law model. We used the \textit{tbabs} absorption model, along with the abundances of the interstellar medium from \citet{Wilms2000}.

To obtain a statistically significant result for the absorption column density, we first fitted the spectrum of a circular region with a radius of $45\,\arcsec$, in the energy range $0.2-10\,$keV. The circle was centered on the region of brightest emission. The result is $\mathrm{N}_\mathrm{H} = (2.25 \pm 0.15)\ $\hcm{23}. Due to the different abundances used in our fit, this value is considerably higher than the one obtained by \citet[][$\mathrm{N}_\mathrm{H} = (1.39 \pm 0.13)\ $\hcm{23}; both values at $90\,\%$ confidence level]{Porquet2003}. When using the same abundances, the values are equal.

\begin{figure}
 \centering
 \includegraphics[clip,width=\linewidth]{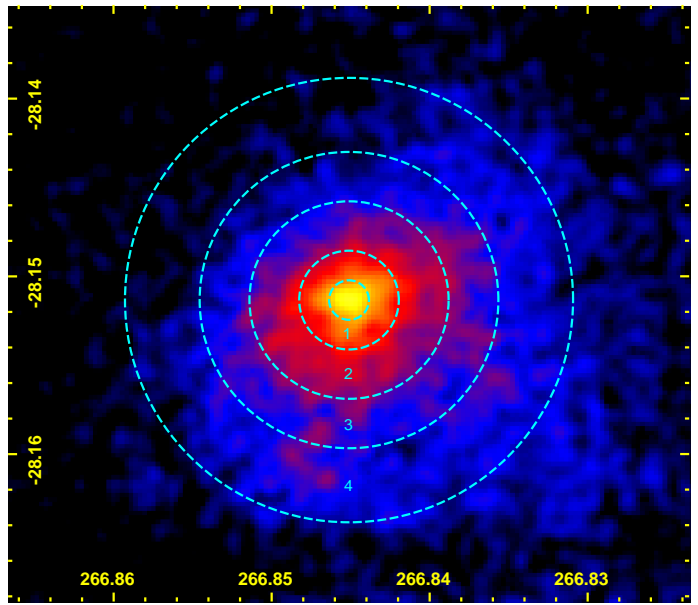}
 \caption{Smoothed \xmm\ count map merged from the MOS and pn data of the 2000 and 2003 observations (see Table~\ref{tab_obs}) in the energy range $3-10\,$keV. The annuli are encompassed by dashed cyan lines (properties listed in Table~\ref{tab_spec}).}
 \label{skymap}
\end{figure}
For the modeling of the emission (see Section~\ref{section_model}), we extracted spectra of four annuli centered on the region of brightest emission. They are illustrated in Fig.~\ref{skymap}, which shows a smoothed \xmm\ count map of \gon, merged from the MOS and pn data of the 2000 and 2003 observations. This map was generated using the \textit{SAOImage ds9} tool \citep{Joye}. 
\begin{table*}
\caption{Results obtained by fitting a power-law spectrum to the \xmm\ data of \gon. All errors are quoted at the $1\,\sigma$ level.}
\label{tab_spec}
\centering
\begin{tabular}{ccccccc}
\hline\hline\noalign{\smallskip}
\multicolumn{1}{c}{Annulus} &
\multicolumn{2}{c}{Radius ($\arcsec$)} &
\multicolumn{1}{c}{$\Gamma$} &
\multicolumn{1}{c}{Unabsorbed flux$^{(*)}$} &
\multicolumn{1}{c}{Surface brightness$^{(*)}$} &
\multicolumn{1}{c}{$\chi^{2}/\mathrm{dof}$}\\
\multicolumn{1}{c}{No.} &
\multicolumn{1}{c}{inner} & 
\multicolumn{1}{c}{outer} &
\multicolumn{1}{c}{} & 
\multicolumn{1}{c}{(\ergcm{-12})} &
\multicolumn{1}{c}{(\ergcm{-16}$\mathrm{arcsec}^{-2}$)} & 
\multicolumn{1}{c}{}\\
\noalign{\smallskip}\hline\noalign{\smallskip}
1 & 4 & 10 & $1.37 \pm 0.09$ & $0.63 \pm 0.02$ & $24.0 \pm 0.6$ & $79/71$\\
2 & 10 & 20 & $1.59 \pm 0.07$ & $1.07 \pm 0.02$ & $11.3 \pm 0.2$ & $97/103$\\
3 & 20 & 30 & $2.01 \pm 0.08$ & $0.91 \pm 0.02$ & $6.8 \pm 0.1$ & $87/81$\\
4 & 30 & 45 & $2.18 \pm 0.08$ & $1.18 \pm 0.02$ & $3.5 \pm 0.1$ & $86/83$\\
\noalign{\smallskip}\hline\noalign{\smallskip}
${}^{(*)}$ Energy range $2.5-10\,$keV
\end{tabular}
\\
\end{table*}
The properties of the annuli are given in Table~\ref{tab_spec}. They are numbered with increasing distance to the center. Since the appearance of \gon\ in X-rays obviously deviates from radial symmetry at least at higher angular distances, we refrained from extracting spectra with greater radii. To increase the statistical significance, we fixed $\mathrm{N}_\mathrm{H}$ to $2.25\,$\hcm{23} for the individual annuli. 
The results obtained from the fitting, as well as the derived surface brightness, are also given in Table~\ref{tab_spec}. According to the reduced $\chi^{2}$ values, a power-law fit yields a very good estimation for each of the spectra. 
\begin{figure}
 \centering
 \resizebox{0.98\hsize}{!}{\begin{turn}{-90}\includegraphics[clip=]{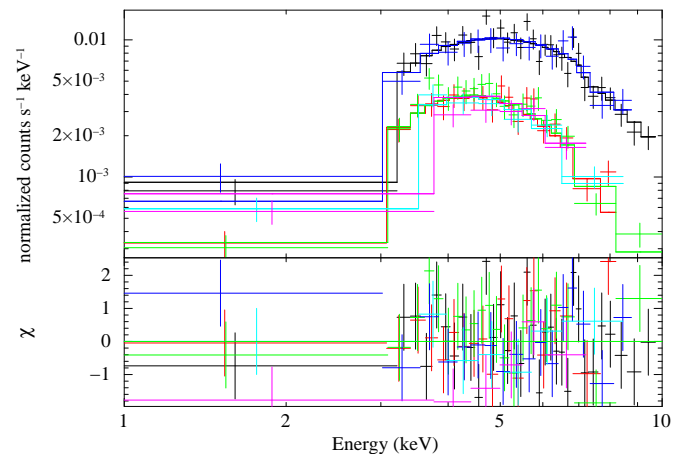}\end{turn}}
 \caption{Energy spectrum extracted from the second annulus. The dark blue and black curves correspond to the pn camera of the 2000 and 2003 observations, respectively. The four lower curves represent MOS1/2 of both observations. The data were fitted in parallel with an absorbed power-law model ({\sc tbabs*cflux*po}).}
 \label{ann2}
\end{figure}
As an example, Fig.~\ref{ann2} shows the spectrum of the second annulus. The strong absorption below $\approx 3\,$keV is clearly visible. 

\section{The model}
\label{section_model}

In the following, we model the observed emission of the PWN in \gon\ with a leptonic scenario. To comprehend the softening of the spectra with greater distance to the pulsar, we adopt a spatially resolved instead of a one-zone model. 

\subsection{Lepton injection spectrum}
\label{injspec}

A pulsar with spin-down luminosity $\dot E$ loses part of its energy in the form of a lepton-dominated particle wind, called the pulsar wind. The leptons of the pulsar wind are accelerated at the wind termination shock and injected into the PWN. The general shape of the injected lepton spectrum is commonly assumed to be a broken power law with the following form \citep[e.g.][]{Kennel1984b,Reynolds1984}:
\begin{equation}
Q_{\mathrm{P}}(E_{\mathrm{e}}) = 
\begin{cases}
\quad Q_{0,\mathrm{P}}\left (\dfrac{E_{\mathrm{e}}}{E_{\mathrm{b,P}}} \right )^{-p_{1}} & \textrm{for} \quad E_{\mathrm{e}} < E_{\mathrm{b,P}} \textrm{,} \smallskip \\
\quad Q_{0,\mathrm{P}}\left (\dfrac{E_{\mathrm{e}}}{E_{\mathrm{b,P}}} \right )^{-p_{2}} & \textrm{for} \quad E_{\mathrm{e}} \ge E_{\mathrm{b,P}} \textrm{,}
\end{cases}
\label{eq_spectrum_inj_pow}
\end{equation}
with the lepton energy $E_{\mathrm{e}}$, the break energy $E_{\mathrm{b,P}}$, the spectral indices $p_{1}$ and $p_{2}$, and the normalization $Q_{0,\mathrm{P}}$ of the spectrum. Since the model focuses on the inner part of the PWN close to the pulsar where mainly young leptons are expected to contribute to the nonthermal emission, we assume that the time scale of our modeling is smaller than any potential variability time scale of the injection spectrum. Furthermore, the applied model only aims to reproduce the measured X- and VHE $\gamma$ radiation, which allows the view to be restricted to the part of the lepton spectrum where $E_{\mathrm{e}}\ge E_{\mathrm{b,P}}$. 

\citet{Spitkovsky2008} suggests an alternative shape for the injected lepton spectrum using two-dimensional particle-in-cell simulations. The best-fit solution to the shape of the lepton population downstream of the termination shock is given by
\begin{equation}
Q_{\mathrm{S}}(E_{\mathrm{e}}) = 
\begin{cases}
Q_{0,\mathrm{S1}}E_{\mathrm{e}}\exp \left( -\frac{E_{\mathrm{e}}}{\Delta E_{\mathrm{1}}} \right) \\ \bigskip
\qquad \qquad \qquad \qquad \qquad \textrm{for} ~ E_{\mathrm{e}} < E_{\mathrm{b,S}} \textrm{,}\\ 
Q_{0,\mathrm{S1}}E_{\mathrm{e}}\exp\left(-\frac{E_{\mathrm{e}}}{\Delta E_{\mathrm{1}}} \right) + Q_{0,\mathrm{S2}}E_{\mathrm{e}}^{-\alpha_{\mathrm{S}}} \\ \bigskip 
\qquad \qquad \qquad \qquad \qquad \textrm{for} ~ E_{\mathrm{b,S}}\le E_{\mathrm{e}} < E_{\mathrm{cut}} \textrm{,}\\
Q_{0,\mathrm{S1}}E_{\mathrm{e}}\exp\left(-\frac{E_{\mathrm{e}}}{\Delta E_{\mathrm{1}}} \right) + Q_{0,\mathrm{S2}}E_{\mathrm{e}}^{-\alpha_{\mathrm{S}}}\exp\left( -\frac{E_{\mathrm{e}}-E_{\mathrm{cut}}}{\Delta E_{\mathrm{cut}}} \right) \\
\qquad \qquad \qquad \qquad \qquad \textrm{for} ~ E_{\mathrm{e}} \ge E_{\mathrm{cut}} \textrm{,}
\end{cases}
\label{eq_spectrum_inj_spit}
\end{equation}
with $\Delta E_{1}$ the width of the relativistic Maxwellian that defines the low-energy domain of the spectrum, $\alpha_{\mathrm{S}}$ the index of the power law, $E_{\mathrm{cut}}$ and $\Delta E_{\mathrm{cut}}$ its cutoff energy and width, and $Q_{0,\mathrm{S1}}$ and $Q_{0,\mathrm{S2}}$ the normalizations of the respective functions. According to \citet{Spitkovsky2008}, $\alpha_{\mathrm{S}}$ lies in the range $2.3-2.5$. The ratios of the relevant quantities obtained from the fit are $E_{\mathrm{b,S}}/\Delta E_{1}\approx 7$, $E_{\mathrm{cut}}/E_{\mathrm{b,S}}\approx 7.5$, and $E_{\mathrm{cut}}/\Delta E_{\mathrm{cut}} \approx 3$. For our modeling, we used the same ratios to preserve the shape of the spectrum. Furthermore, we fixed $\alpha_{\mathrm{S}}$ to a value of $2.4$. 

In the remainder of this paper, we refer to the case of the power-law injection spectrum as \scenone , whereas \scentwo\ corresponds to the injection spectrum from \citet{Spitkovsky2008}. The other parts of the model are identical for both scenarios and are laid out in the following.

The total amount of spin-down power transferred into the energy of the modeled leptons is 
\begin{equation}
  \int_{\eemin}^{E_{\mathrm{e,up}}} Q_{\mathrm{P/S}}(E_{\mathrm{e}})E_{\mathrm{e}}\,\mathrm{d}E_{\mathrm{e}} = \eta \dot E\mathrm{,}
\label{eq_normalization}
\end{equation}
where $\eta$ denotes the conversion efficiency of the pulsar. Using this equation, the normalization of the lepton spectrum can be calculated when the lower and upper integration limits are known. We chose $\eemin = 1\,$erg, which is well above typical break energies of the broken power law ($E_{\mathrm{b,P}}$); however, this value is still low enough not to affect the synchrotron or IC emission in the observed energy ranges, as can be derived from \citet{deJager2009}. Independent of the acceleration mechanism, there are two limits that constrain the maximum energy to which leptons can be accelerated at the termination shock \citep{deJager2009}. The first one states that the gyroradius of the charged particles constituting the pulsar wind must be smaller than the shock radius $R_{\mathrm{S}}$. 
When defining $\epsilon = R_{\mathrm{L}}/R_{\mathrm{S}}$ with $R_{\mathrm{L}}$ the gyroradius, the maximum energy of the leptons is given by \citep{deJager2009}
\begin{equation}
 \eemax = \epsilon e \kappa \sqrt{\dfrac{\sigma}{1 + \sigma}\dfrac{\dot E}{c}}\mathrm{.}
\label{eq_gyro}
\end{equation}
In this equation, $\kappa$ denotes the magnetic compression ratio at the shock and $\sigma$ the magnetization parameter corresponding to the ratio of the magnetic and the particle energy outflow. For want of better estimates and due to the partly strong correlation of the parameters, we fixed $\epsilon = 1$. According to the second limit, a charged particle reaches its maximum energy when the synchrotron losses become as strong as the energy gain. This statement can be rewritten to \citep{deJager2009}
\begin{equation}
 \eemax = 43.7\, B_{\mathrm{S},\mathrm{G}}^{-1/2}\, \mathrm{erg}\mathrm{,}
\end{equation}
with $B_{\mathrm{S},\mathrm{G}}$ the magnetic field strength at the shock in units of G. The lower value of both constraints is then used as $E_{\mathrm{e,up}}$ for the power-law shape or as $E_{\mathrm{cut}}$ when using the spectrum from \citet{Spitkovsky2008}. In the latter case, $E_{\mathrm{e,up}}$ is set to infinity.

\subsection{Outward propagation of the leptons}

The radially symmetric model applied to \gon\ resembles the one introduced by \citet{Schoeck2010}. The lepton plasma is assumed to propagate outwards with a bulk velocity of
\begin{equation}
 v(r) = v_{\mathrm{S}} \left( \dfrac{R_{\mathrm{S}}}{r} \right)^{\alpha}\mathrm{,}
\label{eq_vlep}
\end{equation}   
where $\alpha$ is the index of the adopted power law and $v_{\mathrm{S}}$ the velocity at the wind termination shock. When combining $\kappa$ and $\sigma$ by defining $\xi = \kappa\, \sqrt{\sigma/(1+\sigma)}$, the relation for the magnetic field strength at the shock is given by \citep{Kennel1984a,Sefako2003}:
\begin{equation}
 B_{\mathrm{S}} = \dfrac{\xi}{R_{\mathrm{S}}}\sqrt{\dfrac{\dot E}{c}}\,\mathrm{.}
\label{eq_bsl}
\end{equation} 
Assuming a toroidal magnetic field whose outward propagation is directly connected to that of the leptons, the ideal magnetohydrodynamic (MHD) limit on the assumption of a static system yields \citep{Kennel1984a}
\begin{equation}
 Bvr = B_{\mathrm{S}}v_{\mathrm{S}}R_{\mathrm{S}} = \mathrm{const.} 
\label{eq_mhd}
\end{equation}
As the leptons propagate outwards, they lose energy, leading to a change in the spectral shape. Two fundamental energy loss mechanisms have to be considered: synchrotron radiation of the leptons and adiabatic energy losses. They are given by \citep{deJager1992}
\begin{equation}
 \dfrac{\mathrm{d}E_{\mathrm{e}}}{\mathrm{d}t} = - \dfrac{E_{\mathrm{e}}}{3}\nabla \cdotp \vec{v}_{\bot}(r) - 2.368 \times 10^{-3}(B E_{\mathrm{e}})^2\,\dfrac{\mathrm{erg}}{\mathrm{s}}\,\mathrm{.}
\end{equation}
In this equation, the first term corresponds to adiabatic energy losses and can be calculated using Eq.~\ref{eq_vlep}. The second term denotes the synchrotron losses of the leptons.

\subsection{Implementation and photon emission}

For the numerical implementation of the model, we divided the observed part of the PWN into a large number of concentric subshells in order to simulate the continuous case, exactly as performed by \citet{Schoeck2010}. Using the previously described relations, the lepton injection spectrum is then propagated outwards from one subshell to the next. 

The emitted synchrotron and IC radiation of the corresponding lepton population can be calculated for each subshell. For this, we used the equations given by \citet{Blumenthal1970}. Regarding the IC process, essentially three seed-photon fields are relevant in the case of \gon : the CMB component, IR photons emitted from local dust, and the starlight component. The CMB spectrum is described well by a blackbody distribution for arbitrary locations. As an approximation of the IR and starlight components, we used the interstellar radiation fields of \citet{Porter2005} developed for the GALPROP code \citep{Strong2000}. The emission of the subshells was summed up to shells with the same inner and outer radii as chosen for the annuli of the X-ray analysis.

\subsection{Projection effect}
\label{section_projection}

Since the model is three-dimensional, a shell actually corresponds to a hollow sphere in contrast to the annuli of the X-ray analysis. Such an annulus is merely the two-dimensional projection of the three-dimensional shells. This has two major implications. First, only part of the volume of a shell is visible in the projection represented by an annulus. Second, outer shells add up to an annulus emission. The overall emission from annulus $i$ is given by 
\begin{equation}
 \dfrac{\mathrm{d}N}{\mathrm{d}E}\bigg|_{i,\mathrm{Ann}} = \sum_j \mu_{ij}\,\dfrac{\mathrm{d}N}{\mathrm{d}E}\bigg|_j\mathrm{,} 
\label{eq_3D2D}
\end{equation}
with $\mathrm{d}N/\mathrm{d}E|_j$ the spectrum of shell $j$. This equation includes the assumption that the radiation is emitted isotropically and does not undergo any absorption inside the PWN. The entries of $\mu_{ij}$ can be written in a matrix of size $n\times n$. For the chosen extraction regions from Table~\ref{tab_spec}, $n = 4$ and the matrix can be calculated to 
\begin{equation}
 \mu = \begin{pmatrix}
      0.82 & 0.22 & 0.07 & 0.03 \\ 0 & 0.74 & 0.33 & 0.12 \\ 0 & 0 & 0.59 & 0.26 \\ 0 & 0 & 0 & 0.59
     \end{pmatrix}\mathrm{.}
\end{equation}
This means that, for example, a superposition of $74\,\%$ of the second, $33\,\%$ of the third, and $12\,\%$ of the fourth shell generates the emission of the second annulus. We added up the modeled emission from the shells accordingly in order to obtain the resulting spectra of the annuli.

\subsection{Parameter optimization}
\label{section_par_op}

The previously described model allows us to calculate the synchrotron emission of the annuli for given parameters $p$ (in the power-law case), $R_{\mathrm{S}}$, $v_{\mathrm{S}}$, $\eta$, $\xi$, and $\alpha$. Afterwards the unabsorbed energy flux is computed in six energy bins of equal width ($1.25\,$keV) between $2.5\,$ and $10\,$keV for each annulus. The results are then compared with the measured \xmm\ data by calculating an $X^2$ value as
\begin{equation}
 X^2 = \sum_{i} \left( \dfrac{F_{i,\mathrm{XMM}}-F_{i,\mathrm{mod}}}{\Delta F_{i,\mathrm{XMM}}} \right)^2\mathrm{,}
\end{equation}
where $F_{i,\mathrm{XMM}}$ and $F_{i,\mathrm{mod}}$ denote the measured and modeled flux and $\Delta F_{i,\mathrm{XMM}}$ the statistical 
error in the corresponding energy bin. The parameters of the model are scanned over the allowed range, searching for a minimum of $X^2$. Splitting the spectra into several bins allows optimization of the parameters by taking the overall energy flux into account along with the spectral shape. 

\section{Results of the modeling}
\label{section_results}

The parameter optimization was carried out separately for the aforementioned injection spectra (see Sect.~\ref{injspec}). We assumed a distance of $d=13\,$kpc for both scenarios, following \citet{Camilo2009}. The upper limit on the radius of the termination shock was set to $\varphi_{\mathrm{S}} = 4\,\arcsec$.

Since our model only aims to reproduce the inner part of the PWN, we carried out the parameter optimization while ignoring the $X^2$ values of the fourth annulus.
\begin{figure}
 \centering
 \includegraphics[clip,width=\linewidth]{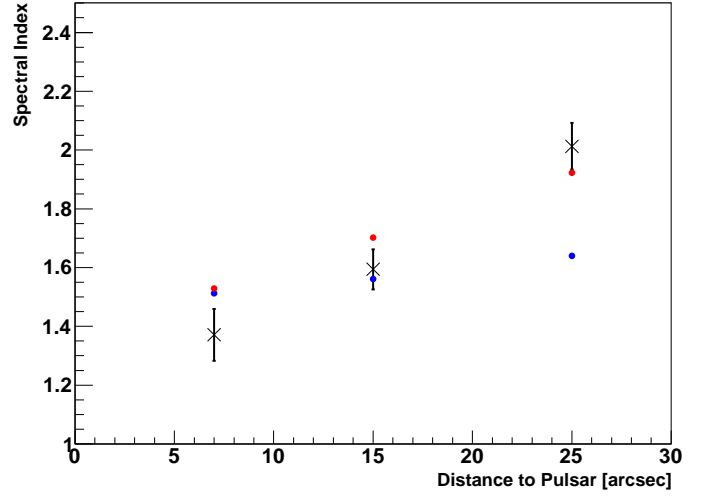}
 \caption{Evolution of the spectral index of the X-ray emission with increasing angular distance to the pulsar. The blue and red points correspond to the results obtained with the optimized parameters of \scenone\ and \scentwo , respectively. The \xmm\ data are shown as black crosses.}
 \label{fig_index_model13}
\end{figure}
The measured and optimized modeled values of the spectral index are presented in Fig.~\ref{fig_index_model13}. Although the modeled index for \scenone\ also increases with growing distance to the pulsar, the slope is too shallow. In \scentwo , by contrast, the spectral index exhibits a stronger steepening and fits the measured data.
\begin{figure}
 \centering
 \includegraphics[clip,width=\linewidth]{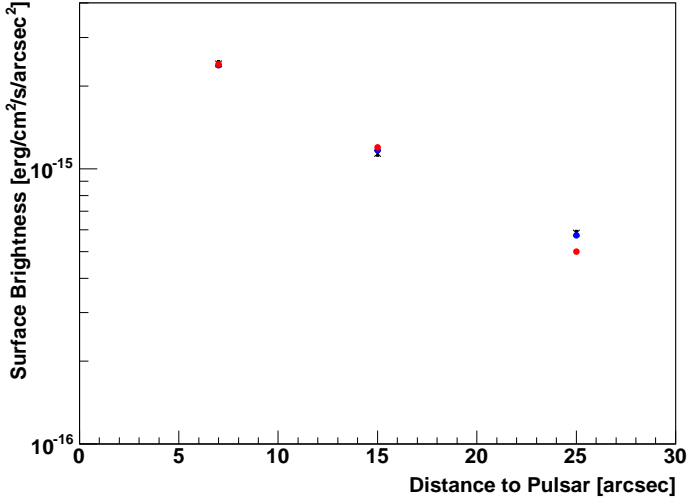}
 \caption{Surface brightness plotted over the distance to the pulsar. The energy range is $2.5-10\,$keV and the blue and red points correspond to \textit{Scenarios I} and \textit{II}, respectively. The black crosses denote the measured \xmm\ data.}
 \label{fig_brightness_model13}
\end{figure}
The evolution of the surface brightness is illustrated in Fig.~\ref{fig_brightness_model13}. The results of the parameter optimization for both scenarios agree with the measured data. As the parameters of the model are partly correlated, one can find several very different parameter sets with similar $X^2$ values. It is thus not possible to designate the fit minimum to a definite parameter set.
\begin{table}
\caption{Lower and upper values of the parameters with an $X^2$ maximum of $15\,\%$ above the 
lowest obtained value for the optimization of the first three annuli.}
\label{tab_resparagyro}
\centering
\begin{tabular}{c|cc}
\hline\hline\noalign{\smallskip}
Parameter &
\scenone &
\scentwo \\
\noalign{\smallskip}\hline\noalign{\smallskip}
$p$ & $1.7$ & $-$ \\
$\xi$ & $0.08$ & $0.22-0.26$ \\
$\alpha$ & $1.00-1.05$ & $1.10-1.15$ \\
$\varphi_{\mathrm{S}}$ [$\arcsec$] & $2.5-4.0$ & $4.0$  \\
$v_{\mathrm{S}}/c$ & $0.13-0.31$ & $0.3$ \\
$\eta$ & $0.4-0.7$ & $0.08$ \\ \hline 
$R_{\mathrm{S}}$ [pc] & $0.16-0.25$ & $0.25$ \\
$B_{\mathrm{S}}$ [$\mu$G] & $3.9-6.2$ & $10.7-12.7$ \\
\hline\noalign{\smallskip}
\end{tabular}
\end{table}
The lower and upper values of the parameters with an $X^2$ maximum of $15\,\%$ above the lowest value for the optimization of the first three annuli are shown in Table~\ref{tab_resparagyro}. However, only certain parameter combinations lead to a good fit result. The best fit results for the conversion efficiency $\eta$ of \scenone\ are higher than the ones of \scentwo , but they lie well within the physical range below $1$ for both scenarios. The velocity at the termination shock $v_{\mathrm{S}}$ is barely constrained for \scenone , but strongly tends to the theoretically expected value of $c/3$ (see e.g. \citet{Kennel1984a}) for \scentwo . $\varphi_{\mathrm{S}}$ remains quite unconstrained for the power-law case, whereas it reaches the aforementioned upper limit of $4\,\arcsec$ when using the injection spectrum of \citet{Spitkovsky2008}. The results for this parameter should, however, not be overemphasized due to its strong correlation with $v_{\mathrm{S}}$. When combining Eqs.~\ref{eq_vlep} and \ref{eq_mhd}, the radial dependence of the magnetic field strength can be calculated for a given parameter set.  
\begin{figure}
 \centering
 \includegraphics[clip,width=\linewidth]{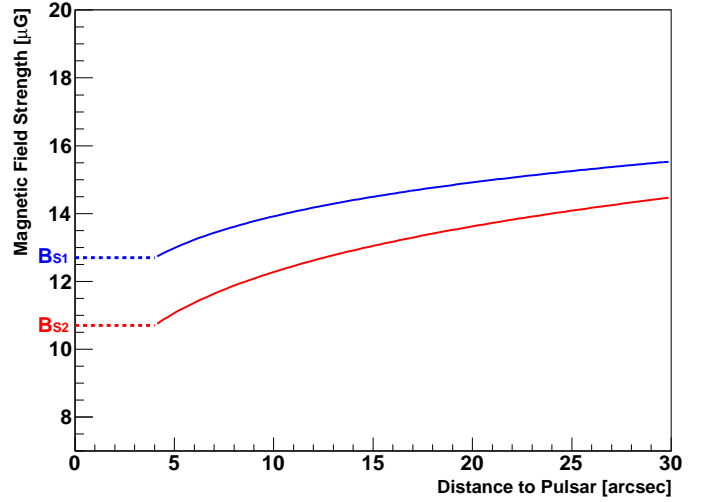}
 \caption{Radial dependence of the magnetic field strength for two parameter sets of \scentwo\ with a low $X^2$ value. The shock magnetic field strength $B_{\mathrm{S}}$ is indicated with a dashed line for both cases.}
 \label{fig_BField}
\end{figure}
This was performed for two sets of \scentwo\ that both exhibit comparably low $X^2$ values but different shock magnetic fields, as shown in Fig.~\ref{fig_BField}. The values of the shock magnetic field strength and velocity index are $B_{\mathrm{S}1}=12.7\,\mu$G and $\alpha_{1}=1.15$ for the first and $B_{\mathrm{S}2}=10.7\,\mu$G and $\alpha_{2}=1.10$ for the second parameter set, respectively. Since the velocity index $\alpha$ is greater than $1$ in both cases, the magnetic field increases with growing distance to the pulsar. However, we are only focusing on the inner part of the PWN. With even greater distance, the ideal MHD limit no longer holds, implying that the magnetic field is not frozen into the particle propagation any longer and might lead to a negative slope farther out. The average magnetic field strength is still rather low for both scenarios, in agreement with the interpretation of \citet{Dubner2008}. Notably the values for \scentwo\ are also in very good agreement with the ones found by \citet[][$B=12-15\,\mu$G]{Tanaka2011}.

\section{Implications for the VHE $\gamma$-ray emission}
\label{section_tev}

As for the synchrotron emission, the IC radiation of the whole modeled area can be calculated for a given parameter set. As already mentioned in Sect.~\ref{introduction}, \gon\ was also detected in the VHE \gr\ regime with H.E.S.S., enabling a comparison of the modeled with the measured data. 
\begin{figure}
 \centering
 \includegraphics[clip,width=\linewidth]{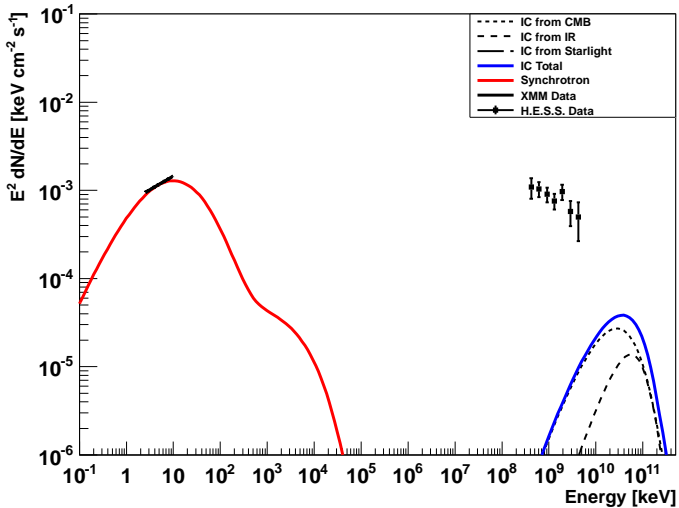}
 \caption{Modeled spectral energy distribution of \gon\ showing the synchrotron and IC emission of the inner $30\,\arcsec$ for \scentwo . Also included are the measured \xmm\ data of the modeled area, as well as the \hess\ data of the whole source \citep{Aharonian2005}.}
 \label{sed}
\end{figure}
The spectral energy distribution with the modeled synchrotron and IC emission for \scentwo\ is shown in Fig.~\ref{sed}. It also comprises the \xmm\ data of the same area, as well as the \hess\ data for the whole source \citep{Aharonian2005}. While the synchrotron emission matches the measured X-ray data well, the IC contribution strongly deviates from the \hess\ data concerning the spectral shape, as well as the flux, implying that the model is not well suited to reproducing the VHE \gr\ emission of this source. This deviation may come from a number of superimposing effects that are discussed in the following. 

Due to the large PSF of \hess\ \citep[$\approx 6\,\arcmin$,][]{Aharonian2006}, \gon\ appears point-like in VHE \gr s. This implies that the modeled and measured VHE $\gamma$ radiation likely originates in regions of different size. Assuming a spatial emission distribution with $\rho \propto \exp(-\theta^2/2\sigma_{\mathrm{source}}^{2})$, \citet{Aharonian2005} obtain an upper limit on the source extent of $\sigma_{\mathrm{source}} < 1.3\,\arcmin$ at $95\,\%$ confidence level. The IC emission within $30\,\arcsec$ (i.e. $0.5\,\arcmin$) around the center, corresponding to the modeled part of the PWN, would contribute only $7\,\%$ to the total VHE $\gamma$ radiation of \gon\ in that case. Furthermore, we only modeled the leptons that are freshly injected at the wind termination shock of the PWN and are confined in the MHD outflow. The lifetime of a lepton scattering photons to the VHE range, however, is longer than the one of a lepton, which emits X-rays in the keV energy range \citep{deJager2009}. Thus a part of the measured \gr s is expected to originate in older leptons that are not energetic enough to emit X-rays any longer. In addition to that, it is also possible that the PWN actually contains a significant amount of energetic hadrons in addition to the leptons. If target material is present, they would produce $\pi^{0}$s, which in turn decay into two \gr s each, increasing the overall VHE $\gamma$ flux of the source. Such a lepto-hadronic scenario has also been suggested for the PWN in the SNR G54.1$+$0.3 by \citet{Li2010}. This source exhibits similar characteristics as \gon\ considering the characteristic age and spin-down luminosity of the pulsar. Finally it has to be noted that the IR and starlight seed-photon fields are only an estimate that does not contain local variations and could furthermore differ from the actual ones due to the uncertain distance to the PWN.

\section{Conclusion}
\label{section_conclusion}

We have presented an extensive analysis of the nonthermal emission of the PWN in the composite SNR \gon. Unlike earlier publications, we were able to better resolve and determine spectral structures in the PWN thanks to an enhanced X-ray data set. Moreover, we performed the first spatially resolved modeling of this source in the X-ray regime and calculated the IC emission of the modeled part of the PWN.

The hardness ratios of two annulus halves, separated by the presumed symmetry axis of the PWN, led us to conclude that we do not see any significant east-west deviation of the spectral properties of the source around the region of brightest emission. Thus we assume that \psrj\ is located in that region. To model the emission, we extracted spectra of four annulus-shaped regions centered on the putative pulsar position. We fixed the absorption column density to the value obtained from a fit of the whole modeled region as this approach significantly reduces the statistical error. The spectral index increases with larger distance to the pulsar, whereas the surface brightness drops.

These characteristics can be explained by assuming a leptonic outflow that suffers synchrotron and adiabatic energy losses as implemented in our model. We carried out the parameter optimization using two different lepton injection spectra, namely a power law (\scenone) and the injection spectrum from \citet{Spitkovsky2008} (\scentwo). Since the adopted model is only valid for the inner part of the PWN where the MHD limit is assumed to hold, we ignored the results of the fourth annulus. The evolution of the surface brightness can be reproduced for both injection spectra. While for \scenone\ the slope of the modeled spectral index is far too low compared to the measured one, with \scentwo\ this characteristic can also be reproduced. Therefore we conclude that this injection spectrum is better suited to explaining the nonthermal emission of this source, however, the energetics of the injection spectrum with the optimized parameters are shifted by about three orders of magnitude to higher values than in other works \citep[see e.g.][]{Slane2011,Fang2010}. The reason for this difference is the positive correlation of the spectral index of the injected lepton population and the one of the resulting synchrotron radiation. The power-law tail of the lepton spectrum (see Eq.~\ref{eq_spectrum_inj_spit}) is sufficient for explaining the averaged spectral index of the whole PWN, as can be seen in these references. In contrast to that, our model is designed to explain the emission of several zones within the PWN, including the innermost one near the termination shock. Since the spectral index of that region is smaller, it is necessary to shift the spectrum to higher energies to get more of a contribution by the leptons of the relativistic Maxwellian.

 With the optimized model parameters it is possible to deduce some of the physical properties of the system. An example of this is the radial dependence of the magnetic field inside the PWN. Interestingly it increases with greater distance according to the model. However, an exact determination of the field strength is not possible since the model parameters are correlated, leading to multiple solutions of parameter sets with a low $X^2$ value.

We calculated the IC emission of the modeled part of the PWN using the parameters optimized to reproduce the X-ray emission. The flux and spectral shape of the modeled radiation do not agree with the measured data. Several possibly concurrent effects may offer an explanation. The unknown extent of \gon\ in VHE \gr s impedes an appropriate comparison of the measured and modeled data. That, in turn, makes it difficult to find out the relative contributions of the other effects, such as TeV emission from hadrons inside the PWN and older leptons that are not accounted for in our modeling. The problem of the unknown extent of \gon\ in VHE \gr s may be solved with future Imaging Atmospheric Cherenkov Telescopes like the Cherenkov Telescope Array (CTA).

\begin{acknowledgements}
We would like to thank J\"{o}rn Wilms for his suggestions concerning the X-ray analysis as well as the editor for the constructive comments that helped to improve the article.
\end{acknowledgements}

\bibliographystyle{aa}
\bibliography{18121}

\begin{thebibliography}{37}
\expandafter\ifx\csname natexlab\endcsname\relax\def\natexlab#1{#1}\fi

\bibitem[{{Aharonian} {et~al.}(2005){Aharonian}, {Akhperjanian}, {Aye},
  {Bazer-Bachi}, {Beilicke}, {Benbow}, {Berge}, {Berghaus}, {Bernl{\"o}hr},
  {Boisson}, {Bolz}, {Borgmeier}, {Braun}, {Breitling}, {Brown}, {Bussons
  Gordo}, {Chadwick}, {Chounet}, {Cornils}, {Costamante}, {Degrange},
  {Djannati-Ata{\"i}}, {O'C.~Drury}, {Dubus}, {Ergin}, {Espigat}, {Feinstein},
  {Fleury}, {Fontaine}, {Funk}, {Gallant}, {Giebels}, {Gillessen}, {Goret},
  {Hadjichristidis}, {Hauser}, {Heinzelmann}, {Henri}, {Hermann}, {Hinton},
  {Hofmann}, {Holleran}, {Horns}, {de Jager}, {Jung}, {Kh{\'e}lifi}, {Komin},
  {Konopelko}, {Latham}, {Le Gallou}, {Lemi{\`e}re}, {Lemoine}, {Leroy},
  {Lohse}, {Marcowith}, {Masterson}, {McComb}, {de Naurois}, {Nolan},
  {Noutsos}, {Orford}, {Osborne}, {Ouchrif}, {Panter}, {Pelletier}, {Pita},
  {P{\"u}hlhofer}, {Punch}, {Raubenheimer}, {Raue}, {Raux}, {Rayner},
  {Redondo}, {Reimer}, {Reimer}, {Ripken}, {Rob}, {Rolland}, {Rowell},
  {Sahakian}, {Saug{\'e}}, {Schlenker}, {Schlickeiser}, {Schuster}, {Schwanke},
  {Siewert}, {Sol}, {Steenkamp}, {Stegmann}, {Tavernet}, {Terrier},
  {Th{\'e}oret}, {Tluczykont}, {Vasileiadis}, {Venter}, {Vincent}, {Visser},
  {V{\"o}lk}, \& {Wagner}}]{Aharonian2005}
{Aharonian}, F., {Akhperjanian}, A.~G., {Aye}, K., {et~al.} 2005, \aap, 432,
  L25

\bibitem[{{Aharonian} {et~al.}(2006){Aharonian}, {Akhperjanian}, {Bazer-Bachi},
  {Beilicke}, {Benbow}, {Berge}, {Bernl{\"o}hr}, {Boisson}, {Bolz}, {Borrel},
  {Braun}, {Breitling}, {Brown}, {B{\"u}hler}, {B{\"u}sching}, {Carrigan},
  {Chadwick}, {Chounet}, {Cornils}, {Costamante}, {Degrange}, {Dickinson},
  {Djannati-Ata{\"i}}, {O'C.~Drury}, {Dubus}, {Egberts}, {Emmanoulopoulos},
  {Espigat}, {Feinstein}, {Ferrero}, {Fiasson}, {Fontaine}, {Funk}, {Funk},
  {Gallant}, {Giebels}, {Glicenstein}, {Goret}, {Hadjichristidis}, {Hauser},
  {Hauser}, {Heinzelmann}, {Henri}, {Hermann}, {Hinton}, {Hofmann}, {Holleran},
  {Horns}, {Jacholkowska}, {de Jager}, {Kh{\'e}lifi}, {Komin}, {Konopelko},
  {Kosack}, {Latham}, {Le Gallou}, {Lemi{\`e}re}, {Lemoine-Goumard}, {Lohse},
  {Martin}, {Martineau-Huynh}, {Marcowith}, {Masterson}, {McComb}, {de
  Naurois}, {Nedbal}, {Nolan}, {Noutsos}, {Orford}, {Osborne}, {Ouchrif},
  {Panter}, {Pelletier}, {Pita}, {P{\"u}hlhofer}, {Punch}, {Raubenheimer},
  {Raue}, {Rayner}, {Reimer}, {Reimer}, {Ripken}, {Rob}, {Rolland}, {Rowell},
  {Sahakian}, {Saug{\'e}}, {Schlenker}, {Schlickeiser}, {Schwanke}, {Sol},
  {Spangler}, {Spanier}, {Steenkamp}, {Stegmann}, {Superina}, {Tavernet},
  {Terrier}, {Th{\'e}oret}, {Tluczykont}, {van Eldik}, {Vasileiadis}, {Venter},
  {Vincent}, {V{\"o}lk}, {Wagner}, \& {Ward}}]{Aharonian2006}
{Aharonian}, F., {Akhperjanian}, A.~G., {Bazer-Bachi}, A.~R., {et~al.} 2006,
  \aap, 457, 899

\bibitem[{{Arnaud}(1996)}]{Arnaud1996}
{Arnaud}, K.~A. 1996, in Astronomical Society of the Pacific Conference Series,
  Vol. 101, Astronomical Data Analysis Software and Systems V, ed.
  {G.~H.~Jacoby \& J.~Barnes}, 17

\bibitem[{{Arnaud} {et~al.}(2001){Arnaud}, {Neumann}, {Aghanim}, {Gastaud},
  {Majerowicz}, \& {Hughes}}]{Arnaud2001}
{Arnaud}, M., {Neumann}, D.~M., {Aghanim}, N., {et~al.} 2001, \aap, 365, L80

\bibitem[{{Blackburn}(1995)}]{Blackburn1995}
{Blackburn}, J.~K. 1995, in Astronomical Society of the Pacific Conference
  Series, Vol.~77, Astronomical Data Analysis Software and Systems IV, ed.
  {R.~A.~Shaw, H.~E.~Payne, \& J.~J.~E.~Hayes}, 367

\bibitem[{{Blumenthal} \& {Gould}(1970)}]{Blumenthal1970}
{Blumenthal}, G.~R. \& {Gould}, R.~J. 1970, Reviews of Modern Physics, 42, 237

\bibitem[{{Camilo} {et~al.}(2009){Camilo}, {Ransom}, {Gaensler}, \&
  {Lorimer}}]{Camilo2009}
{Camilo}, F., {Ransom}, S.~M., {Gaensler}, B.~M., \& {Lorimer}, D.~R. 2009,
  \apjl, 700, L34

\bibitem[{{Cordes} \& {Lazio}(2002)}]{Cordes2002}
{Cordes}, J.~M. \& {Lazio}, T.~J.~W. 2002, ArXiv Astrophysics e-prints

\bibitem[{{de Jager} \& {Djannati-Ata{\"i}}(2009)}]{deJager2009}
{de Jager}, O.~C. \& {Djannati-Ata{\"i}}, A. 2009, in Astrophysics and Space
  Science Library, ed. {W.~Becker}, Vol. 357, 451

\bibitem[{{de Jager} \& {Harding}(1992)}]{deJager1992}
{de Jager}, O.~C. \& {Harding}, A.~K. 1992, \apj, 396, 161

\bibitem[{{Dubner} {et~al.}(2008){Dubner}, {Giacani}, \&
  {Decourchelle}}]{Dubner2008}
{Dubner}, G., {Giacani}, E., \& {Decourchelle}, A. 2008, \aap, 487, 1033

\bibitem[{{Fang} \& {Zhang}(2010)}]{Fang2010}
{Fang}, J. \& {Zhang}, L. 2010, \aap, 515, A20

\bibitem[{{Gaensler} {et~al.}(2001){Gaensler}, {Pivovaroff}, \&
  {Garmire}}]{Gaensler2001}
{Gaensler}, B.~M., {Pivovaroff}, M.~J., \& {Garmire}, G.~P. 2001, \apjl, 556,
  L107

\bibitem[{{Gaensler} \& {Slane}(2006)}]{Gaensler2006}
{Gaensler}, B.~M. \& {Slane}, P.~O. 2006, \araa, 44, 17

\bibitem[{{Helfand} \& {Becker}(1987)}]{Helfand1987}
{Helfand}, D.~J. \& {Becker}, R.~H. 1987, \apj, 314, 203

\bibitem[{{Jansen} {et~al.}(2001){Jansen}, {Lumb}, {Altieri}, {Clavel}, {Ehle},
  {Erd}, {Gabriel}, {Guainazzi}, {Gondoin}, {Much}, {Munoz}, {Santos},
  {Schartel}, {Texier}, \& {Vacanti}}]{Jansen2001}
{Jansen}, F., {Lumb}, D., {Altieri}, B., {et~al.} 2001, \aap, 365, L1

\bibitem[{{Joye} \& {Mandel}(2003)}]{Joye}
{Joye}, W.~A. \& {Mandel}, E. 2003, in Astronomical Society of the Pacific
  Conference Series, Vol. 295, Astronomical Data Analysis Software and Systems
  XII, ed. {H.~E.~Payne, R.~I.~Jedrzejewski, \& R.~N.~Hook}, 489

\bibitem[{{Kennel} \& {Coroniti}(1984{\natexlab{a}})}]{Kennel1984a}
{Kennel}, C.~F. \& {Coroniti}, F.~V. 1984{\natexlab{a}}, \apj, 283, 694

\bibitem[{{Kennel} \& {Coroniti}(1984{\natexlab{b}})}]{Kennel1984b}
{Kennel}, C.~F. \& {Coroniti}, F.~V. 1984{\natexlab{b}}, \apj, 283, 710

\bibitem[{{Kesteven}(1968)}]{Kesteven1968}
{Kesteven}, M.~J.~L. 1968, Australian Journal of Physics, 21, 369

\bibitem[{{Li} {et~al.}(2010){Li}, {Chen}, \& {Zhang}}]{Li2010}
{Li}, H., {Chen}, Y., \& {Zhang}, L. 2010, \mnras, 408, L80

\bibitem[{{Majerowicz} {et~al.}(2002){Majerowicz}, {Neumann}, \&
  {Reiprich}}]{Majerowicz2002}
{Majerowicz}, S., {Neumann}, D.~M., \& {Reiprich}, T.~H. 2002, \aap, 394, 77

\bibitem[{{Manchester} \& {Taylor}(1977)}]{Manchester1977}
{Manchester}, R.~N. \& {Taylor}, J.~H. 1977, {Pulsars.}, ed. {Smith, F.~G.}

\bibitem[{{Mereghetti} {et~al.}(1998){Mereghetti}, {Sidoli}, \&
  {Israel}}]{Mereghetti1998}
{Mereghetti}, S., {Sidoli}, L., \& {Israel}, G.~L. 1998, \aap, 331, L77

\bibitem[{{Park} {et~al.}(2006){Park}, {Kashyap}, {Siemiginowska}, {van Dyk},
  {Zezas}, {Heinke}, \& {Wargelin}}]{Park2006}
{Park}, T., {Kashyap}, V.~L., {Siemiginowska}, A., {et~al.} 2006, \apj, 652,
  610

\bibitem[{{Porquet} {et~al.}(2003){Porquet}, {Decourchelle}, \&
  {Warwick}}]{Porquet2003}
{Porquet}, D., {Decourchelle}, A., \& {Warwick}, R.~S. 2003, \aap, 401, 197

\bibitem[{{Porter} \& {et al.}(2005)}]{Porter2005}
{Porter}, T.~A. \& {et al.} 2005, in International Cosmic Ray Conference,
  Vol.~4, International Cosmic Ray Conference, 77

\bibitem[{{Reynolds} \& {Chevalier}(1984)}]{Reynolds1984}
{Reynolds}, S.~P. \& {Chevalier}, R.~A. 1984, \apj, 278, 630

\bibitem[{{Sch{\"o}ck} {et~al.}(2010){Sch{\"o}ck}, {B{\"u}sching}, {de Jager},
  {Eger}, \& {Vorster}}]{Schoeck2010}
{Sch{\"o}ck}, F.~M., {B{\"u}sching}, I., {de Jager}, O.~C., {Eger}, P., \&
  {Vorster}, M.~J. 2010, \aap, 515, A109

\bibitem[{{Sefako} \& {de Jager}(2003)}]{Sefako2003}
{Sefako}, R.~R. \& {de Jager}, O.~C. 2003, \apj, 593, 1013

\bibitem[{{Slane}(2011)}]{Slane2011}
{Slane}, P. 2011, in High-Energy Emission from Pulsars and their Systems, ed.
  {D.~F.~Torres \& N.~Rea}, 373

\bibitem[{{Spitkovsky}(2008)}]{Spitkovsky2008}
{Spitkovsky}, A. 2008, \apjl, 682, L5

\bibitem[{{Strong} {et~al.}(2000){Strong}, {Moskalenko}, \&
  {Reimer}}]{Strong2000}
{Strong}, A.~W., {Moskalenko}, I.~V., \& {Reimer}, O. 2000, \apj, 537, 763

\bibitem[{{Str{\"u}der} {et~al.}(2001){Str{\"u}der}, {Briel}, {Dennerl},
  {Hartmann}, {Kendziorra}, {Meidinger}, {Pfeffermann}, {Reppin}, {Aschenbach},
  {Bornemann}, {Br{\"a}uninger}, {Burkert}, {Elender}, {Freyberg}, {Haberl},
  {Hartner}, {Heuschmann}, {Hippmann}, {Kastelic}, {Kemmer}, {Kettenring},
  {Kink}, {Krause}, {M{\"u}ller}, {Oppitz}, {Pietsch}, {Popp}, {Predehl},
  {Read}, {Stephan}, {St{\"o}tter}, {Tr{\"u}mper}, {Holl}, {Kemmer}, {Soltau},
  {St{\"o}tter}, {Weber}, {Weichert}, {von Zanthier}, {Carathanassis}, {Lutz},
  {Richter}, {Solc}, {B{\"o}ttcher}, {Kuster}, {Staubert}, {Abbey}, {Holland},
  {Turner}, {Balasini}, {Bignami}, {La Palombara}, {Villa}, {Buttler},
  {Gianini}, {Lain{\'e}}, {Lumb}, \& {Dhez}}]{Strueder2001}
{Str{\"u}der}, L., {Briel}, U., {Dennerl}, K., {et~al.} 2001, \aap, 365, L18

\bibitem[{{Tanaka} \& {Takahara}(2011)}]{Tanaka2011}
{Tanaka}, S.~J. \& {Takahara}, F. 2011, \apj, 741, 40

\bibitem[{{Turner} {et~al.}(2001){Turner}, {Abbey}, {Arnaud}, {Balasini},
  {Barbera}, {Belsole}, {Bennie}, {Bernard}, {Bignami}, {Boer}, {Briel},
  {Butler}, {Cara}, {Chabaud}, {Cole}, {Collura}, {Conte}, {Cros}, {Denby},
  {Dhez}, {Di Coco}, {Dowson}, {Ferrando}, {Ghizzardi}, {Gianotti}, {Goodall},
  {Gretton}, {Griffiths}, {Hainaut}, {Hochedez}, {Holland}, {Jourdain},
  {Kendziorra}, {Lagostina}, {Laine}, {La Palombara}, {Lortholary}, {Lumb},
  {Marty}, {Molendi}, {Pigot}, {Poindron}, {Pounds}, {Reeves}, {Reppin},
  {Rothenflug}, {Salvetat}, {Sauvageot}, {Schmitt}, {Sembay}, {Short},
  {Spragg}, {Stephen}, {Str{\"u}der}, {Tiengo}, {Trifoglio}, {Tr{\"u}mper},
  {Vercellone}, {Vigroux}, {Villa}, {Ward}, {Whitehead}, \&
  {Zonca}}]{Turner2001}
{Turner}, M.~J.~L., {Abbey}, A., {Arnaud}, M., {et~al.} 2001, \aap, 365, L27

\bibitem[{{Wilms} {et~al.}(2000){Wilms}, {Allen}, \& {McCray}}]{Wilms2000}
{Wilms}, J., {Allen}, A., \& {McCray}, R. 2000, \apj, 542, 914

\end{thebibliography}

\end{document}